\documentclass[reprint,superscriptaddress,amsmath,amssymb,aps,prc]
{revtex4-1}

\usepackage{graphicx}
\usepackage{hyperref}
\usepackage{cleveref}
\usepackage[english]{babel}

\begin{document}

\title{An alternative approach to populate and study the $^{229}$Th nuclear clock isomer}

    \author{M. Verlinde}
    \email{Matthias.Verlinde@kuleuven.be}
    \affiliation{Institute for Nuclear and Radiation Physics, KU Leuven, B-3001 Leuven, Belgium}
	\author{S. Kraemer}
    \affiliation{Institute for Nuclear and Radiation Physics, KU Leuven, B-3001 Leuven, Belgium}
    \author{J. Moens}
    \affiliation{Institute for Nuclear and Radiation Physics, KU Leuven, B-3001 Leuven, Belgium}
    \author{K. Chrysaldis}
    \affiliation{CERN, CH-1211 Geneva 23, Switzerland}
    \author{J.G. Correia}
    \affiliation{C2TN, DECN, Instituto Superior T\'{e}cnico, Universidade de Lisboa, 2695-066 Bobadela, Portugal}
    \author{S. Cottenier}
    \affiliation{Center for Molecular Modeling, Ghent University, 9052 Zwijnaarde, Belgium}
    \affiliation{Department of Electrical Energy, Metals, Mechanical Constructions and Systems, Ghent University, 9052 Zwijnaarde, Belgium}
    \author{H. De Witte}
    \affiliation{Institute for Nuclear and Radiation Physics, KU Leuven, B-3001 Leuven, Belgium}
    \author{D.V. Fedorov}
    \affiliation{Petersburg Nuclear Physics Institute, NRC Kurchatov Institute, Gatchina 188300, Russia}
    \author{V.N Fedosseev}
    \affiliation{CERN, CH-1211 Geneva 23, Switzerland}
    \author{R. Ferrer}
    \affiliation{Institute for Nuclear and Radiation Physics, KU Leuven, B-3001 Leuven, Belgium}
    \author{L.M. Fraile}
    \affiliation{Grupo F\'{i}sica Nuclear IPARCOS, Universidad Complutense de Madrid, E-28040 Madrid, Spain}
    \author{S. Geldhof}
    \affiliation{Department of Physics, University of Jyv\"{a}skyl\"{a}, FI-40014 Jyv\"{a}skyl\"{a}, Finland}
    \author{C.A. Granados}
    \affiliation{CERN, CH-1211 Geneva 23, Switzerland}
    \author{M. Laatiaoui}
    \affiliation{Helmholtz-Institut Mainz, D-55128 Mainz, Germany}
    \affiliation{Institute f\"{u}r Kernchemie, Universit\"{a}t Mainz, D-55128 Mainz, Germany}
    \author{T.A.L. Lima}
    \affiliation{Institute for Nuclear and Radiation Physics, KU Leuven, B-3001 Leuven, Belgium}
    \author{P-C Lin}
    \affiliation{Institute for Nuclear and Radiation Physics, KU Leuven, B-3001 Leuven, Belgium}
    \author{V. Manea}
    \affiliation{Institute for Nuclear and Radiation Physics, KU Leuven, B-3001 Leuven, Belgium}
    \author{B.A. Marsh}
    \affiliation{CERN, CH-1211 Geneva 23, Switzerland}
    \author{I. Moore}
    \affiliation{Department of Physics, University of Jyv\"{a}skyl\"{a}, FI-40014 Jyv\"{a}skyl\"{a}, Finland}
    \author{L.M.C. Pereira}
    \affiliation{Institute for Nuclear and Radiation Physics, KU Leuven, B-3001 Leuven, Belgium}
    \author{S. Raeder}
    \affiliation{Helmholtz-Institut Mainz, D-55128 Mainz, Germany}
    \affiliation{GSI Helmholtzzentrum f\"{u}r Schwerionenforschung GmbH, D-64291 Darmstadt, Germany}
    \author{P. Van den Bergh}
    \affiliation{Institute for Nuclear and Radiation Physics, KU Leuven, B-3001 Leuven, Belgium}
    \author{P. Van Duppen}
    \affiliation{Institute for Nuclear and Radiation Physics, KU Leuven, B-3001 Leuven, Belgium}
    \author{A. Vantomme}
    \affiliation{Institute for Nuclear and Radiation Physics, KU Leuven, B-3001 Leuven, Belgium}
    \author{E. Verstraelen}
    \affiliation{Institute for Nuclear and Radiation Physics, KU Leuven, B-3001 Leuven, Belgium}
    \author{U. Wahl}
    \affiliation{Institute for Nuclear and Radiation Physics, KU Leuven, B-3001 Leuven, Belgium}
    \affiliation{C2TN, DECN, Instituto Superior T\'{e}cnico, Universidade de Lisboa, 2695-066 Bobadela, Portugal}
    \author{S.G. Wilkins}
    \affiliation{CERN, CH-1211 Geneva 23, Switzerland}


\begin{abstract}
\noindent
\textbf{A new approach to observe the radiative decay of the $^{229}$Th nuclear isomer, and to determine its energy and radiative lifetime, is presented. Situated at a uniquely low excitation energy, this nuclear state might be a key ingredient for the development of a nuclear clock, a nuclear laser and the search for time variations of the fundamental constants. The isomer's $\gamma$ decay towards the ground state will be studied with a high-resolution VUV spectrometer after its production by the $\beta$ decay of $^{229}$Ac. The novel production method presents a number of advantages asserting its competitive nature with respect to the commonly used $^{233}$U $\alpha$-decay recoil source. In this paper, a feasibility analysis of this new concept, and an experimental investigation of its key ingredients, using a pure $^{229}$Ac ion beam produced at the ISOLDE radioactive beam facility, is reported.}

\end{abstract}

\maketitle

\section{Introduction} \label{sec:intro}
\noindent
In 1976, experimental evidence was reported pointing to a low-lying nuclear state in the $^{229}$Th isotope at an excitation energy, E$_{\textrm{iso}}$, below $100$ eV \cite{Kroger1976}. Subsequent experiments lowered the predicted excitation energy of this $^{229}$Th isomer down to the eV range, which is unique on the nuclear chart \cite{Burke2008,Gulda2002,Canty1977,Bemis1988,Barci2003,Ruchowska2006,Reich1990,Helmer1994,Irwin1997}. This exeptionally low excitation energy of the isomer initiated numerous research projects aimed at unveiling its characteristics. In the last decade, breakthrough experiments offered a proof of its existence, studied the isomer's nuclear moments and provided confirmation of an excitation energy in the VUV part of the spectrum (E$_{\textrm{iso}} < 18.3$ eV) \cite{Beck2007,Beck2009,Wense2016,Thielking2018}. The available experimental data on the isomer in $^{229}$Th are summarized in \cref{fig:decayscheme}. The main feature of this isomer, its low excitation energy, promises a range of future applications which rely on the extension of laser manipulation techniques from the electron shell towards the nuclear domain, more specifically: a direct VUV laser excitation of a nuclear transition could be within reach for the first time. In addition to the creation of a nuclear clock, matching or even exceeding the performance of present day atomic clocks, this isomeric nuclear state could be decisive in the development of the first nuclear laser and more fundamentally; in the study of the time dependence of fundamental constants of nature \cite{Peik2003,Peik2015,Campbell2012,Berengut2009,Tkalya2011}.\\
\medskip
\\Working towards a laser excitation of a nuclear transition, fulfilling the isomer's potential, requires, however, improvement in the precision and accuracy of both excitation-energy and unknown radiative-lifetime observables. The present-day accepted value places the isomer's excitation energy at E$_{\textrm{iso}} = 7.8(5)$ eV. This value was obtained through indirect high-resolution $\gamma$-spectroscopy data of inter- and intraband transitions in ground- and isomeric state collective bands \cite{Beck2007}. The large experimental uncertainty of 0.5 eV (corresponding to $\approx 10$ nm or $\approx 120$ THz) in combination with an expected narrow nuclear resonance of $10^{-4}$ Hz, presents a practical stumbling block for the direct excitation of this nuclear transition using coherent light sources. Moreover, recent studies shed doubt on the accuracy of the current excitation energy \cite{Tkalya2015}. To improve both the accuracy and the precision of the isomer's key observables, a variety of physics techniques are implemented, with one branch focusing on a direct measurement of the electromagnetic deexcitation of $^{229\textrm{m}}$Th embedded in a host material.\\
\medskip
\\To study and detect the VUV photons originating from the $^{229\textrm{m}}$Th-$^{229}$Th transition, the $\alpha$ decay of $^{233}$U (T$_{1/2} = 1.6 \cdot 10^5$ years) is frequently used to populate the isomer \cite{Beck2007,Wense2016,Zhao2012,Stellmer2016}. With a probability of $\approx$ 2\% per $\alpha$ decay, $^{233}$U decays to $^{229\textrm{m}}$Th, whose 84 keV recoil energy is thermalized in a gas or solid host after which its decay products can be scrutinized. Due to the unusually low excitation energy, the electromagnetic environment where the isomer is finally situated, determines its decay mode. In case the thorium electrons have a free quantum state available at an energy equal to E$_{\textrm{iso}}$ above its current state, satisfying conservation of energy, the isomer's dominant decay channel is internal conversion (IC). A half-life of T$_{1/2} = 7 \pm 1$ $\mu$s was reported for this decay channel when the isomer was deposited on the surface of a Multi-Channel Plate (MCP) detector \cite{Wense2016}. In the other case, nuclear $\gamma$ decay will persist, albeit with a much longer lifetime, $\alpha_{IC}=\frac{T_{1/2,\gamma}}{T_{1/2,IC}} \approx 10^9$ \cite{Seiferle2017a}. This rare feature allows a certain freedom in tuning the environment to select the favored decay mode. However, at the same time, it requires sufficient understanding of the $^{229\textrm{m}}$Th electromagnetic environment.\\
\medskip
\\In this paper, a novel approach to observe the radiative decay and to measure the radiative energy and lifetime of $^{229\textrm{m}}$Th is described. It is based on an alternative production method, whereby the isomer is populated via the $\beta$ decay of $^{229}$Ac, produced online. VUV spectroscopy is then performed after implantation of $^{229}$Ac in a large band gap crystal (E$_{bandgap}>$ E$_{\textrm{iso}}$, to block the IC channel), where it decays to $^{229\textrm{m}}$Th. In the next section, a detailed comparison between the aforementioned production methods will be presented, in line with the goals of the proposed experiment. The feasibility of the proposed experiment will then be outlined in section \ref{sec:vuv}. Three preparatory experiments have been conducted and reported in sections \ref{sec:isolde}-\ref{sec:branching}-\ref{sec:ic}-\ref{sec:channel} together with some preliminary results. First, the ability to produce an intense beams of actinium ions was verified at ISOLDE, CERN, for the first time (see section \ref{sec:isolde}). Second, as a key performance indicator of $^{229}$Ac for producing the isomer, an attempt was made to measure the true total feeding probability, $^{229}$Ac$-^{229\textrm{m}}$Th (see section \ref{sec:branching}-\ref{sec:ic}). Finally, the lattice positions of the $^{229}$Ac atoms, after implantation in a CaF$_2$ host, were studied (see section \ref{sec:channel}).

\section{The $\beta$ decay of $^{229}$A\lowercase{c}}\label{sec:229Ac}
\begin{figure}[!htbp]
\begin{center}
\includegraphics[width=\columnwidth]{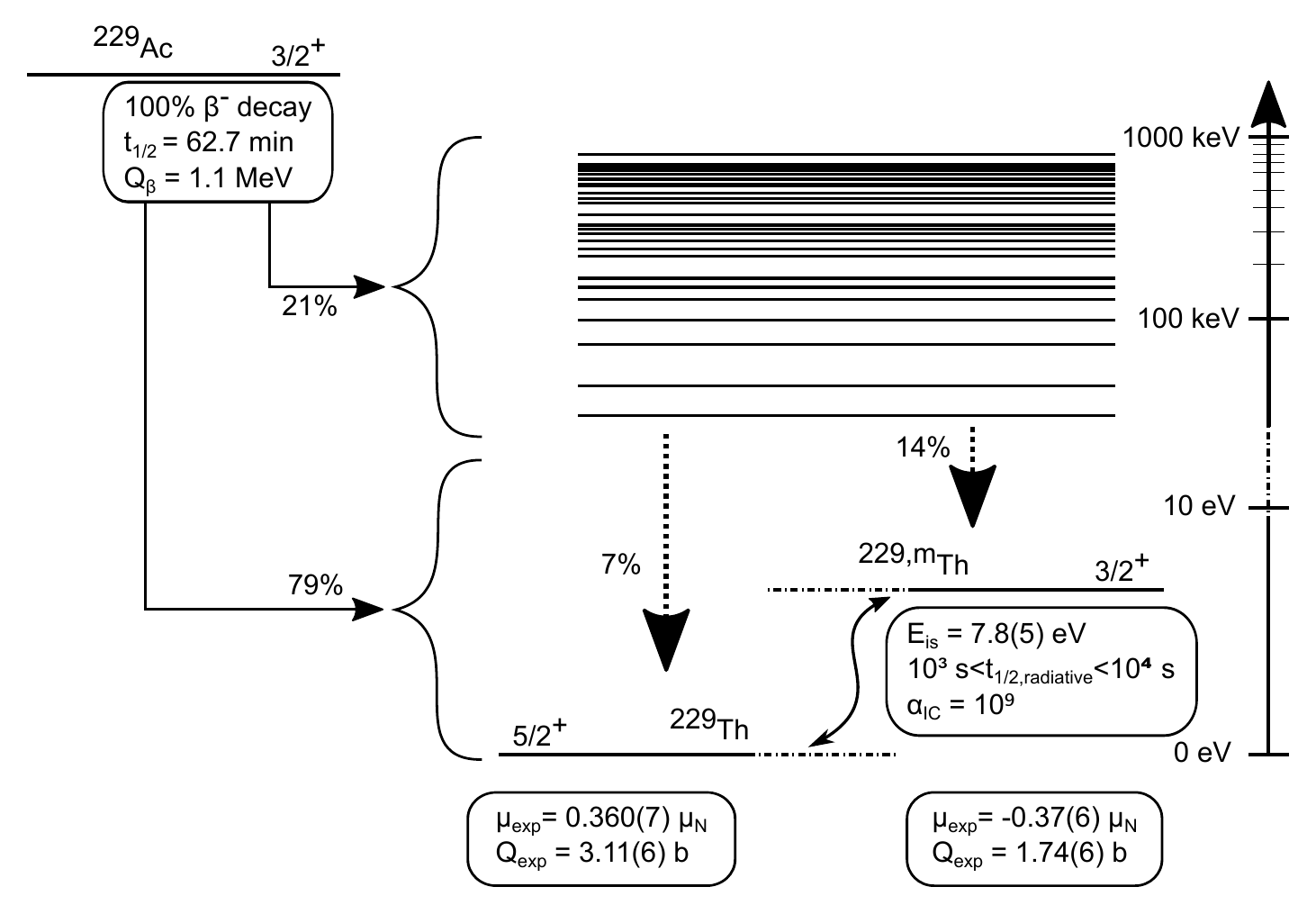}
\caption{The known experimental data on the isomer, $^{229\textrm{m}}$Th, mentioned in section \ref{sec:intro}, are shown alongside the most important parameters concerning the $\beta$ decay of $^{229}$Ac, mentioned in section \ref{sec:229Ac}. The total feeding probabilities of the $^{229}$Ac $\beta^-$ decay to both isomer and ground state are shown. The value for the radiative half-life of the isomer, T$_{1/2,\textrm{radiative}}$, is a theoretical prediction. The $\mu_{\textrm{exp}}$ and Q$_{\textrm{exp}}$ are the experimental magnetic- and quadrupole moments, respectively \cite{Thielking2018,Campbell2011,Seiferle2017a,NNDC,Minkov2017a,Tkalya2015}.
}\label{fig:decayscheme}
\end{center}
\end{figure}

\noindent
To produce $^{229\textrm{m}}$Th via radioactive decay, three methods exist: $\alpha$ decay from the $^{233}$U (T$_{1/2} = 1.6 \cdot 10^5$ years) mother nucleus, electron capture decay from $^{229}$Pa (T$_{1/2} = 1.5$ days) and $\beta$ decay from $^{229}$Ac (T$_{1/2}$ = 62.7 min). The $^{233}$U source has been used extensively as means to produce the isomer, also when aiming at a direct decay measurement of the low-energy photons. However, some aspects of this production method can be brought forward as possible reasons for the absence of conclusive results. The $^{233}$U radioactive decay produces energetic $\alpha$ particles ($\approx 4.8$ MeV). The background signal that these particles induce through radioluminescence in the host structure, is a strong competitor for the isomeric decay signal, which is, additionally, hampered by the small $\approx$ 2\% branching ratio \cite{VonderWense2018}. Next, absorbing the $84$ keV recoil energy of the $^{229\textrm{m}}$Th particles in the host material will cause local pockets of lattice damage close to the unknown point of arrival. As the large bandgap of the host crystal is essential for the radiative decay of the isomer to happen, this structural damage and the unknown stopping point of the isomer after implantation could induce new levels in the band gap, restricting the ability of the host to suppress the IC channel. \\
\medskip
\\As an alternative to $^{233}$U, $^{229}$Pa and $^{229}$Ac exist. Both isotopes suffer from low production yields, although renewed interest can be found in $^{229}$Pa. Only recently, an efficient way of producing $^{229}$Ac, online, was reported \cite{Ferrer2017}. It enables to study $^{229\textrm{m}}$Th with intense beams of $^{229}$Ac, for which the $\beta$ decay presents a number of key advantages in comparison to the $^{233}$U $\alpha$ decay. A summary of the decay characteristics of $^{229}$Ac is given in \cref{fig:decayscheme}. The advantages can be listed as follows:

\begin{itemize}
\item The online production in a UC$_x$ target via a laser ion source at ISOLDE, CERN, using a proven ionization scheme, will provide an intense and pure source of $^{229}$Ac (section \ref{sec:isolde}).
\item From the point of the feeding probability, the literature branching ratio $^{229}\textrm{Ac}-^{229\textrm{m}}\textrm{Th}$ $\geq$ 14$\%$ anticipates an increase in the signal to background ratio by at least a factor of seven compared to $^{233}$U (section \ref{sec:branching}) \cite{NNDC}.
\item The $\beta$ decay of $^{229}$Ac provides an almost recoil free (E$_{\textrm{recoil}}$ $\approx$ 2 eV) $^{229\textrm{m}}$Th production, locking $^{229\textrm{m}}$Th in the place of implantation, establishing an unchanged band gap.
\item The $\beta$ particles, emerging after the $^{229}$Ac decay, are likely to produce a significantly lower radioluminescence in comparison with the high stopping power $\alpha$ particles and the recoiling $^{229}$Th nucleus, both emerging after $^{233}$U decay.
\item The $^{229}$Ac half-life (T$_{1/2}$ = 62.7 min) opens a time window for annealing the host crystal after implantation, before radiation detection takes place (see section \ref{sec:channel}) \cite{Pereira2011}.
\end{itemize}
The advantages of using $^{229}$Ac $\beta$ decay over $^{233}$U $\alpha$ decay are explained in detail below. First, an indirect $\beta$-decay branching ratio of at least 14\% towards the isomer in the $^{229}$Th daughter has been experimentally determined \cite{NNDC}. This presents only a lower limit for the feeding probability towards the isomer, because, per $\beta$ decay, a probability of 79\% exists for the decay path to go towards the $^{229}$Th isomer or ground state directly, indistinguishable in these data. The feeding probability should, therefore, lie in the region 14\% $<\lambda_{\beta,\textrm{isomer}}<$ 93\%. An experiment to determine this probability will be outlined in section \ref{sec:branching}. Second, after the $^{229}$Ac $\beta$ decay, the $^{229}$Th nuclei experience a maximum recoil energy of only $2.3$ eV \cite{Kofoed-Hansen1948}, which is around an order of magnitude smaller than the typical displacement energies in a crystal lattice . Therefore, it is expected that after implantation, both $^{229}$Ac and its daughter $^{229}$Th reside in the same position on the crystal lattice. This assumption drastically simplifies an accurate understanding of the electromagnetic environment of the isomer, as the 2.3 eV recoil energy does not induce any additional damage to the lattice. An experiment, aiming for an accurate determination of the isomer's position in the host's lattice, will be discussed in section \ref{sec:channel}. Third, when implanting $^{229}$Ac in a crystal host, the emitted $\beta$ particles will result in a lower radiatively-induced background. The $\beta$ particle stopping power is around three orders of magnitude lower than that for $\alpha$'s and much lower than that of the recoiling nuclei. Finally, in case the aforementioned experiment shows an unfavorable lattice location of the isomeric nuclei, one where E$_{\textrm{bandgap}}>$ E$_{\textrm{iso}}$ is not fulfilled, the lifetime of the $^{229}$Ac isotope allows for host annealing to be done after implantation. In this way, the probability of a favorable lattice site occupation can be increased \cite{Pereira2011}. In combination, these numbers present an increase of at least a factor seven of the signal-to-noise ratio in comparison to the conventional $^{233}$U production method, potentially further enlarged by the unknown total feeding probability, as well as better control of the lattice occupation and a reduction of the radioluminescence background signal that will be quantified in the next section.

\section{Quantitative analysis of the VUV spectroscopy concept}\label{sec:vuv}
\noindent
The proposed experiment to study $^{229\textrm{m}}$Th, produced online via $^{229}$Ac, will consist of two main components. First of all, the 30 keV $^{229}$Ac ion beam will be implanted in the bulk of a thin large-bandgap host crystal. The reduced thickness of the sample will minimize the background (see below), while, more importantly, the large band gap inhibits the IC decay channel of the isomer. Multiple options for the host crystal exist, such as CaF$_2$, MgF$_2$, Na$_2$ThF$_6$, LiCaAlF$_6$, LiSrAlF$_6$, YLIF$_4$ and more exotic, frozen noble gases \cite{Hehlen2013,VanderHeyden1987}. For the purpose of this paper, CaF$_2$ is selected, presenting a band gap of 11.6-12.1 eV \cite{Dessovic2014}. $^{229}$Ac ions implanted in CaF$_2$ at 30 keV, have a projected range of 18.2 nm with a straggling of 3.5 nm. Depending on the outcome of the implantation studies of $^{229}$Ac in CaF$_2$ (see section \ref{sec:channel}), the implantation period of around two $^{229}$Ac half-lives ($\approx$ 2 hours), can be followed by \textit{in situ} annealing procedures. In the second part of the experiment, the crystal will be moved to the spectrometer port, where the low energy photons of the $^{229\textrm{m}}$Th-$^{229}$Th transition will be studied by a VUV spectrometer (e.g. Resonance VM92). This device delivers a spectral resolution better than 1 nm with a 150 $\mu$m slit, standard grating of 1800 grooves/mm, f$\#$ number of 2.2 and a grating efficiency $>$ 15$\%$ in combination with tailored entrance optics, cooled electronics and an MCP-based detection setup. A spectral resolution of 0.1 nm is within reach after decreasing the slit's dimensions, at the cost of efficiency. During detection, a second foil will be irradiated with the $^{229}$Ac beam simultaneously to minimize duty cycle losses of the online beam time.\\
\medskip
\\In order to quantify the expected signal strength at the $^{229\textrm{m}}$Th-$^{229}$Th photon energy, a detailed feasibility analysis can be done. Based on theoretical cross sections for 1.4 GeV protons impinging on a $^{238}$UC$_x$ target at the ISOLDE facility of CERN, a $^{229}$Ac ion beam intensity of 10$^7$ pps (see section \ref{sec:isolde}) is estimated. In the aforementioned measuring cycle with a detection time of $\pm$3 h, and an assumed 1 h radiative half-life of the isomer, approximately 3$\cdot$10$^5$ VUV photons s$^{-1}$ are to be expected from the isomeric decay, considering a conservative feeding of only 14$\%$ (this number represents an average over the 3 h of detection time). This number can be compared to the expected photon rates from $^{229\textrm{m}}$Th ions, produced by $^{233}$U, recoiling from a thin foil ($\approx$ 150 s$^{-1}$), from a 1 cm$^3$ $^{233}$U-doped CaF$_2$ crystal (4$\cdot$10$^4$ s$^{-1}$) or using synchrotron radiation to excite $^{229}$Th-doped crystals (10$^6$ s$^{-1}$) \cite{Hehlen2013}. The total VUV detection efficiency considers the geometrical efficiency of the entrance slit, the acceptance solid angle of the spectrometer, the efficiency of the grating and the quantum efficiency of the CCD camera setup, adding to a total efficiency of 0.003\%. Taking this number into account, the isomer decay signal should correspond to a total of around 9 cps situated in the range 159(10) nm, see \cref{tab:parameters}. The most significant background-signal contributions, detected in the MCP detector, are divided in three categories. First, the detection of primary radiation from the decay of $^{229}$Ac and its daughters is minimized due to passive radiation shielding and the limited measuring time on each implantation foil. Second, the radioluminescence of the decay products of $^{229}$Ac is to be considered. When fully stopped, the 5 MeV $\alpha$ particles emerging from the decay of $^{229}$Th release a total of around 10$^4$ photons. By limiting the CaF$_2$ layer to a thickness of 50 nm, the $\alpha$ particles will only lose around 10 keV leading to a total detection rate smaller than 10$^{\textrm{-}4}$ cps at the detector during the decay cycle, predominantly in the spectral region above 200 nm. A similar story holds for the signal due to radioluminescence from the $\beta$ particles, which, in the region below 200 nm, is predominantly owing to Cerenkov radiation. Only losing about 10 eV in the 50 nm implantation foil, this specific background rate should be below $10^{\textrm{-}4}$ cps at the detector site \cite{Stellmer2016,Stellmer2015}. Finally, the ISOLDE facility provides the opportunity to run the experiment with other actinium isotopes. In combination with the possibility to check the obtained signal with the lasers of the ISOLDE ion source on/off (see section \ref{sec:isolde}), these extra measures should deliver key advantages in characterizing the background signal in a reproducible way.

\begin{table}[!htbp]
    \begin{tabular}{l|r}
    \hline
    \hline
    Description & Estimate \\
    \hline
    $^{229}$Ac beam intensity (pps) & $1.0 \cdot 10^7$  \\
    $^{229}$Ac atoms implanted in 2h & $7.2\cdot 10^{10}$ \\
    $^{229}$Ac atoms after 2h implantation& $4.0 \cdot 10^{10}$ \\
    Isomeric decays in 3h & $3.4\cdot 10^{9}$ \\
    VUV photon rate (cps)  & $9.3\cdot 10^{0}$ \\
    Radioluminescence background & $<10^{-4}$  \\ 
    ($^{229}$Ac $\beta$ decay) (cps)& \\    
    Radioluminescence background & $<10^{-4}$  \\
    ($^{229}$Th $\alpha$ decay) (cps)& \\
    \hline
    \hline
    \end{tabular}
    \caption{Theoretical estimates for the expected count rate (cps) at the excitation energy of the isomer in the VUV spectrometer detector for a specific $^{229}$Ac production rate (pps), compared with the anticipated background contributions.}\label{tab:parameters}
\end{table}

\section{The production of $^{229}$A\lowercase{c} at ISOLDE, CERN} \label{sec:isolde}
\begin{figure}[!htbp]
\begin{center}
\includegraphics[width=0.5\columnwidth]{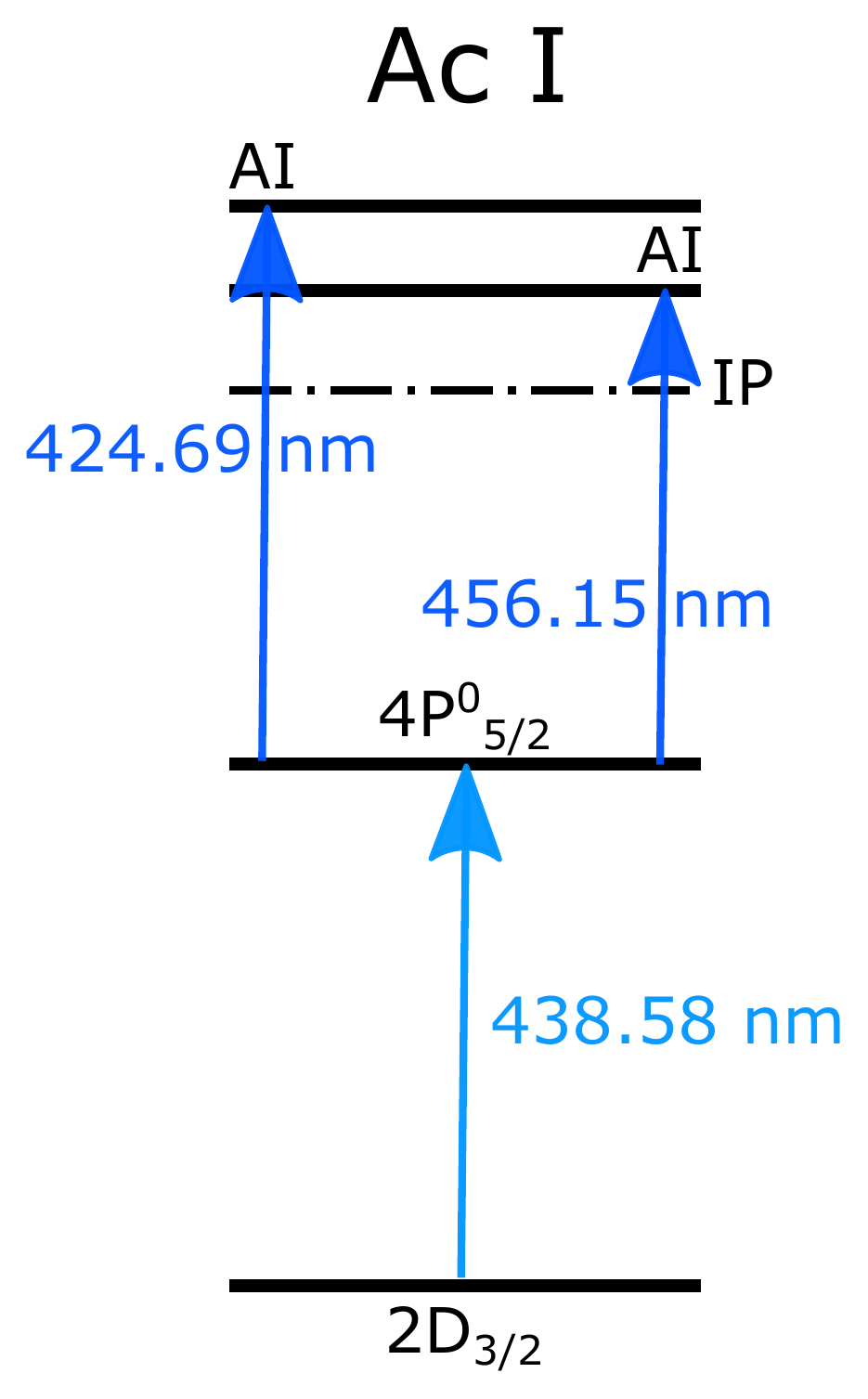}
\caption{The two laser-ionization schemes used for the first production of laser-ionized actinium ions at the RILIS laser ion source of ISOLDE, CERN  \cite{Raeder2013,Ferrer2017}. The ionization efficiency for both auto-ionizing levels is similar. }\label{fig:Acionizationscheme}
\end{center}{}
\end{figure}

\noindent
To allow for a high intensity, pure $^{229}$Ac source, the described experiments take place at the ISOLDE facility in CERN, Geneva \cite{Borge2017}. In this radioactive ion beam facility, protons at an energy of 1.4 GeV, and an average current of 2 $\mu$A, produced by CERN’s PS-Booster proton accelerator, irradiate a thick ($\approx$100 g/cm$^2$) $^{238}$UC$_x$ target heated to about 2000$^{\circ}$C. The nuclear reaction products which arise from these energetic collisions are extracted from a hot target, predominantly in a neutral atomic state, via diffusion and effusion into an ionization cavity. Subsequently, they are ionized, mainly via laser resonant ionization \cite{Fedosseev2017}, increasing the total yield and beam purity significantly. After laser ionization, the ions are extracted and accelerated to 30 kV, mass separated and sent to different experiments. Recently, a number of efficient laser ionization schemes for actinium isotopes are developed \cite{Raeder2013,Ferrer2017}. The two laser schemes used to produce actinium ions for the experiments discussed in sections \ref{sec:branching} and \ref{sec:channel} are shown in \cref{fig:Acionizationscheme}. Ionization with either 424.69 nm or 456.15 nm revealed similar ionization efficiency, resulting in a laser-ionized $^{227}$Ac production rate of about $4\cdot10^7$ pps as measured at a Faraday cup in the focal plane of the separator. Based on a calculated cross section for the $^{229}$Ac isotope, 1.2 $\cdot 10^8$ particles/$\mu$A are produced in a typical $^{238}$UC$_x$ target (the yield could be increased by an order of magnitude when using a $^{232}$Th target \cite{Armbruster2004}). Within conservative estimates, the laser ion source at ISOLDE, aiming for the known Ac laser ionization scheme of \cref{fig:Acionizationscheme}, should be capable of producing $^{229}$Ac beams of around 10$^7$ pps, delivered in a spot of diameter 4 mm. Due to suboptimal ion transport conditions, caused by technical problems in the ISOLDE separator, a production rate of about 10$^6$ particles s$^{-1}$ was observed at the branching ratio setup (discussed in section \ref{sec:branching}), for the specific isotope $^{229}$Ac. Through the combination of the known Ac ionization scheme and the ISOLDE facility, it is possible to produce isotopically-pure, high-intensity beams of $^{229}$Ac via laser resonant ionization techniques in an online facility, opening the door for the exploitation of the $^{229}$Ac isotope as a viable source of the isomer in $^{229}$Th.

\section{Branching ratio measurement}\label{sec:branching}
\begin{figure*}[!htbp]
\resizebox{1\textwidth}{!}{
\includegraphics{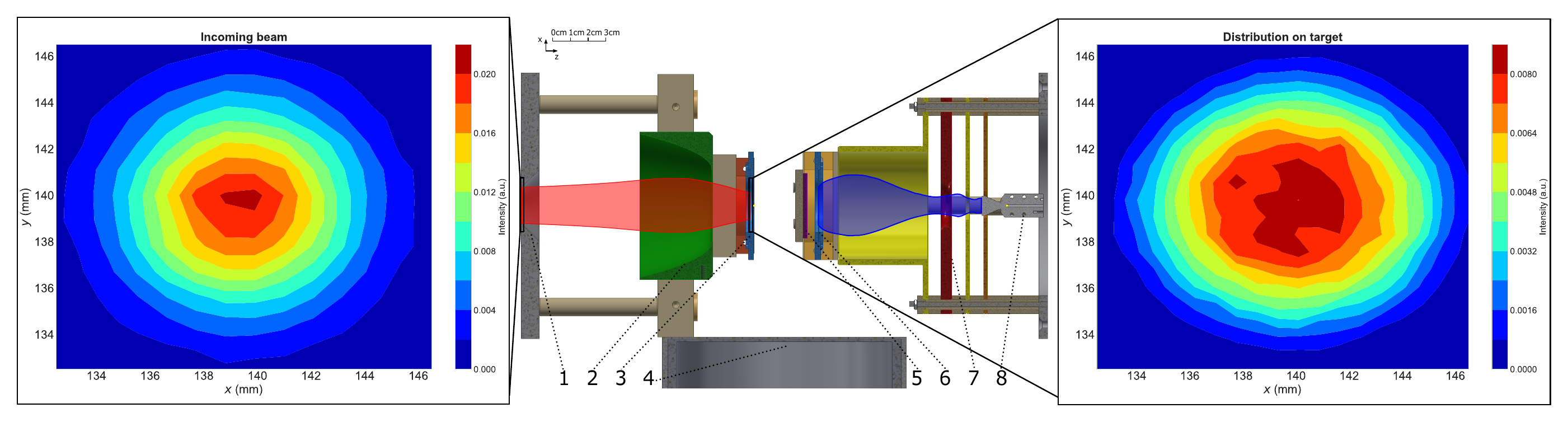}
}
\caption{A schematic view of the setup for the measurement of the branching ratio with diaphragm (1), deceleration electrode (2) and target in implantation position (3), the HPGe detector (4), $\beta$ detector (5), target  in detection position (6), acceleration electrode for the low-energy electrons (7) and channeltron detector (8). The beam envelope (depicted in red) and the expected distributions at the diaphragm and the target (inset on the left and right, respectively) for typical beam characteristics are shown. The blue envelope corresponds to the escaping conversion electrons leaving the target foil for detection in the channeltron.}
\label{fig:softlanding}       
\end{figure*}

\noindent
One key feature of the new approach to study the ${}^{229}$Th isomer is its expected strong feeding in the decay of ${}^{229}$Ac. To measure the unknown total branching ratio towards the isomer, a dedicated setup detecting the isomeric decay via its low-energy IC electrons in time-delayed coincidence with the $\beta$ decay of the actinium mother nucleus has been developed, see \cref{fig:softlanding}. Differentiating the low-energy internal conversion electrons stemming from the decay of the isomer from other low-energy electrons (e.g. $\beta$ and $\gamma$ radiation-induced secondary electrons and conversion electrons from higher lying states in ${}^{229}\textrm{Th}$) is based on the time-dependence of the signal. Where the former shows an exponential decay with a half-life of $7$ $\mu$s, the latter is of prompt nature within a time window of 500 ns. The experimentally determined half-life of $7$ $\mu$s, however, is expected to be dependent on the chemical environment of the isomer, which will be different for the present experiment. The setup in \cref{fig:softlanding} consists of two components. First, the ${}^{229}\textrm{Ac}^{1+}$ radioactive ion beam will be implanted in a suitable host material. This host is chosen such that after $\beta$-decay the chemical environment around the thorium dopant favors the IC decay channel of the isomer over the radiative one in order to take advantage of the shorter lifetime ($\alpha_{IC}\approx 10^9$) of this channel and the efficient detection of low-energy electrons \cite{Seiferle2017a}. Internal conversion will take place if the electromagnetic environment allows the excitation of surrounding electrons with the isomeric energy E$_{\textrm{iso}}$. This condition is easily fulfilled in metallic hosts, where a continuous distribution of electronic quantum states, referred to as Density of States (DOS), without any intermittent bandgap is available. Finally, the setup consists of a rotatable target holder with a thin metallic foil (25 $\mu$m) as implantation host material, which can be lifted, turned and positioned from the low-energy implantation site to a detection site where the time-delayed coincidences will be performed (see central part of \cref{fig:softlanding}).\\
\medskip
\\In order for the low-energy IC electrons to escape from the host material, the actinium dopant must be implanted close to the metal/host surface. To accomplish this, the highly energetic actinium ions, which are delivered to the implantation section of the setup, first encounter a deceleration electrode of parabolic shape which can decelerate the radioactive ion beam to a few hundreds of eV and focus it on the target to obtain a shallow implantation profile (see left and right \cref{fig:softlanding}). The design has been optimized for an efficient deposition of actinium within a spot of 20 mm diameter on the target, using the SIMION electrostatic solver and ion trajectory simulator \cite{Dahl2000}. A typical beam with 30$\pi$ mm mrad emittance and a Gaussian spatial distribution of 7 mm full-width-at-half-maximum at 30 keV energy passes a diaphragm with 20 mm inner diameter at the entrance to the setup. The voltage of the deceleration electrode (typically 1-3 kV below the target voltage) is tuned to obtain optimal focusing of the decelerated beam on the target. The simulation shows that with this conservative estimate of typical beam parameters at ISOLDE and a correctly aligned beam, efficiencies of 95$\%$ or higher at implantation energies of 100 eV are obtained. The magnitude of the deceleration electrode's dimensions and the large target size make the design less sensitive to deviations of the incoming actinium ion beam with respect to the central axis. Higher implantation energies decrease this sensitivity even further but reduce at the same time the detection efficiency of low-energy conversion electrons stemming from the isomeric decay as outlined in the next section.\\
\medskip
\\After an implantation period of approximately one half-life, the target is rotated towards the detection section. $\beta$ radiation emitted in the decay of ${}^{229}\textrm{Ac}$ is registered by a passivated implanted planar silicon detector (PIPS, Mirion 30X30500eb). $\gamma$ radiation can be detected using a 70$\%$ relative efficiency high-purity germanium detector (Canberra HPGe 88045). IC electrons leaving the bulk of the target are accelerated by a 4 kV electrode and guided towards a channeltron detector (Sjuts KBL10RS). The germanium, silicon and channeltron detector signals are recorded on an event-by-event basis together with an absolute time stamp and the signals of time-to-amplitude converters. This time information will be used to discriminate between signals in the channeltron from secondary electrons produced by $\beta$- and $\gamma$ radiation and direct $\beta$ radiation. Secondary electrons created by $\beta$- and $\gamma$ radiation are expected to produce a prompt timing signal in the electron detector, while the conversion electrons from isomeric decay will be detected as time-delayed coincidences, characterized by the $\mu s$ lifetime of the isomer, with the direct $\beta$ particles. Time-delayed coincidences between the low-energy electron-, $\beta$- and $\gamma$-signals allow to determine the direct and indirect isomeric state feeding, respectively. The conversion electron detector is designed to accept, at high efficiency, electrons that emerge from the target with a $2\pi$ solid angle distribution with a maximum energy of 8 eV. Because real IC electrons will show a lower maximum kinetic energy due to the work function of the metal, they will be more swiftly accelerated towards the channeltron upon leaving the target host and, thus, detected with an optimal efficiency. The design has been verified and optimized using electron trajectory simulations. IC electrons emerging from the target within a round spot of radius $10$ mm are detected with a $>99\%$ efficiency, allowing for a large implantation spot size on the target. The high-purity germanium detector has an estimated detection efficiency of $0.8\%$ at an energy of 569 keV, while the PIPS detector accepts betas with a $40 \%$ geometrical efficiency. The total efficiency for detecting the low-energy electrons from the IC process are discussed in the next section.\\
\medskip
\\The setup has been used in a test experiment at ISOLDE and the functionality of the newly developed components was verified. Two different targets, consisting of a thin niobium and gold layer grown on a Mylar film, have been used for implantation at 30, 5 and 2 keV. The choice of the host material and the implantation energies is motivated in the next section. Time spectra, created by gating on known $\gamma$-ray energies, feeding directly or indirectly the isomer, and on the low-energy electron detector signal, were investigated. Using the known ${}^{229}\textrm{Ac}$ decay scheme and the 6$\%$ total low-electron detection efficiency for 2 keV implantation in Nb (see next section), it can be concluded, with a 95\% confidence level, that no 7 $\mu$s $\gamma$-electron decay signal was observed. Possible scenario’s to explain this might be a different (shorter) half-life for internal conversion of the isomer, embedded in a metallic host material, or a different $\beta$-decay feeding pattern of the isomer as compared to literature.

\section{Internal Conversion in a Metal}\label{sec:ic}
\begin{figure*}[!htbp]
\resizebox{0.99\textwidth}{!}{
\includegraphics{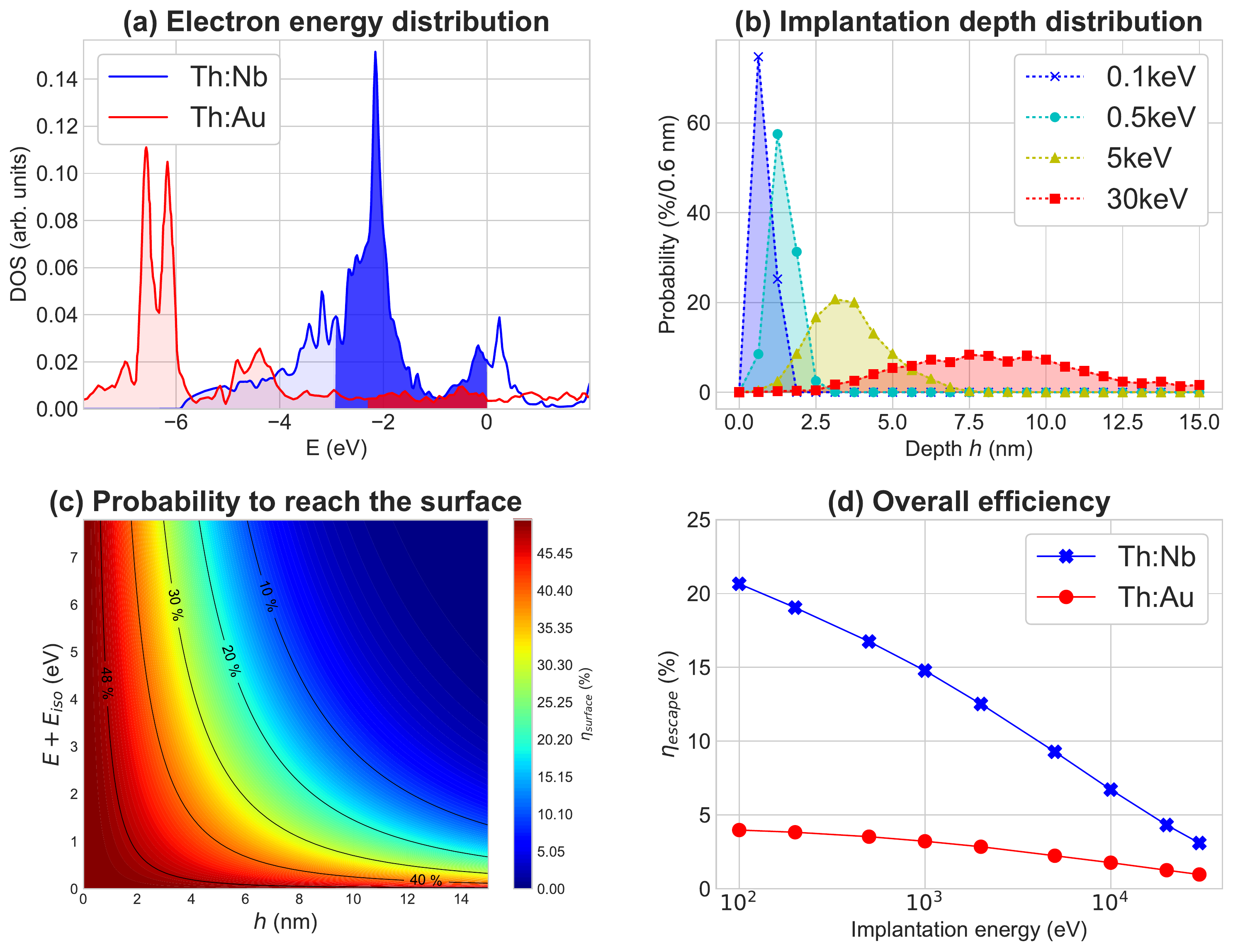}
}
\caption{Simulation of the escape probability of the electron from the host metal: The conversion electron DOS is shown in (a) for Th:Nb and Th:Au, obtained form DFT calculations. E$_{\textrm{F}}=0$, therefore, the shaded area corresponds to occupied states. The Light shaded areas represent electrons that are energetically not allowed to leave the bulk while the dark shaded areas represent electrons that - if not scattered - will cross the surface. The work functions for gold and niobium are 5.5 and 4.9 eV, repectively. (b) depicts the implantation depth distribution in niobium for different implantation energies, calculated via SRIM software. (c) shows the probability of an electron to reach the bulk surface as a function of its implantation depth and its energy with respect to the Fermi surface. The estimated escape probability given by \cref{escape} as a function of implantation energy is given in (d) for niobium and gold hosts. For details see text.}
\label{fig:2}       
\end{figure*}

\noindent
In order to estimate the escape probability for the low-energy IC electrons of interest, internal conversion is modeled in the bulk material, neglecting the surface effects of the first few monolayers. First, the energy of the isomeric decay E$_\textrm{iso}$ is transferred from the nucleus to an electron in an occupied bound state of the valence band if a suitable unoccupied electronic state is available. Next, this electron travels towards the surface of the bulk material while losing energy in scattering processes. Finally, upon arriving at the surface, the electron is released if it overcomes the surface potential barrier. Below, these three processes are treated independently.\\
\medskip
\\In the present model, the converted electrons are characterized by their energy, E, expressed relative to the Fermi energy of the metal E$_\textrm{F}=0$. The occupied levels are, thus, characterized by $E<0$. The energy distribution of valence-band quantum states is given by the local density of states at the dopant's location and is calculated using density functional theory (DFT) \cite{Lejaeghere2016,Cottenier}. A projection on hydrogen s-wave functions selects the components with zero angular momentum, whereas higher angular momenta are neglected as their relative spatial overlap with the nucleus compared to s-states is negligible and, therefore, less significant in the internal conversion decay process. At $0$ K, quantum states are occupied by electrons up to the Fermi energy level E$_\textrm{F}$, higher energetic states remain unoccupied. Electrons, excited by the internal conversion process, move from an initially occupied state at energy E to an unoccupied final state at an energy E$_\textrm{f}$ $= \textrm{E} + \textrm{E}_\textrm{iso}$. The expected effect of the internal conversion decay on the distribution of electron energies, obtained from DFT calculations, is shown for gold and niobium in \cref{fig:2} (a).
\noindent
 After excitation via internal conversion decay of the nucleus, a conversion electron travels from its initial position at depth h in the bulk material towards the surface and scatters elastically and inelastically. Inelastic scattering is characterized by the inelastic mean free path $\lambda_\textrm{IMFP}$, which is a function of energy and can be approximated as universal for elementary metals \cite{Seah1979}. The starting position is defined by the implantation energy. $^{229}$Ac ions are implanted into the host at low energies of a few hundreds to a few thousands eV. At $100$ eV implantation energy the mean implantation depth is around $1$ nm for niobium with a lattice constant of $0.330$ nm \cite{Smirnov1966}. Typical implantation profiles were obtained with the binary collision approximation using the SRIM software package and are shown in \cref{fig:2} (b) \cite{Ziegler2004}. The probability of conversion electrons emitted isotropically at depth h with energy E, with respect to the Fermi level, to reach the surface is given by:
\begin{equation}
\begin{split}
&\eta_\textrm{surface} \left( \textrm{E},\textrm{h} \right)=\\
&\frac{1}{\pi}	\int_0^{\pi/2}  \textrm{exp}\left(\frac{\textrm{h}}{\textrm{cos}\left(\theta\right) \lambda_\textrm{IMFP}\left(\textrm{E}+\textrm{E}_\textrm{iso}\right)}\right)\textrm{sin}(\theta)d\theta,
\end{split}
\end{equation}
with $\theta$ equal to the polar angle. This probability is shown in \cref{fig:2} (c) for elementary metals. At the surface the electrons have to overcome the surface potential characterized by the metal’s work function, E$_\textrm{WF}$. Electrons stemming from electronic states close to the Fermi energy gain enough energy during internal conversion, while electrons from the lower part of the DOS distribution remain bound in the bulk material. Only electrons with an energy  E + E$_{\textrm{iso}}$ $ \geq \textrm{E}_\textrm{WF}$ can leave the bulk. The escape probability thus becomes:
\begin{equation}
\begin{split}
&\eta_\textrm{escape} \left(h\right) = \\
&\int_{-\textrm{E}_\textrm{iso}+\textrm{E}_\textrm{WF}}^{0} \frac{ \eta_\textrm{surface}\left(\textrm{E},\textrm{h}\right) \textrm{DOS}\left(\textrm{E}\right)d\textrm{E}
}{\int^{0}_{- \textrm{E}_\textrm{iso} } \textrm{DOS}\left( \textrm{E} \right)d\textrm{E}}.
\label{escape}
\end{split}
\end{equation}
Consequently, a high efficiency is obtained for host materials with a low work function and a high density of states in the region between -E$_\textrm{iso}$+E$_{WF}$ and the Fermi level of the DOS, as shown  by the dark shaded regions in \cref{fig:2} (a). Electrons stemming from the lower energy region of the energy distribution (light shaded in the figure) remain bound and are not detected. Low-energy implantation deposits actinium ions close to the surface, such that electrons can reach the boundary without being scattered but deep enough that distortions by the surface are avoided.
\\
\medskip
\\
DFT calculations were used to search for a host material with a DOS that maximizes the number of available electrons capable of reaching the surface. From these simulations, $\eta_\textrm{escape}$ was computed as shown in \cref{tab:escape}, after implantation at 100 eV energy. The choice of a suitable host material for the branching ratio experiment not only depends on the electronic structure, but also on the contamination of the material at the surface and in the bulk. The presence of contaminants in the bulk can change the electron kinetic energy distribution significantly, while an oxide layer at the surface additionally alters the work function. This oxide layer has to be thin enough, such that electrons can pass through it without experiencing losses and such that the workfunction is not significantly altered. A sample material with high purity should be chosen in order to be able to neglect their influence. Gold, as an inert element, is not exhibiting an oxide layer and is, additionally, available with high purity. However, at 100 eV implantation energy, its escape probability is limited to $4\%$. Based on these efficiencies and as a trade-off between a favorable electronic structure, a low expected oxidation rate and the availability of pure sample material, niobium has been chosen as additional target material. The expected escape efficiencies at different implantation energies for gold and niobium are shown in \cref{fig:2} (d).

\begin{table}[!htbp]
    \begin{tabular}{l|r||r|r}
    \hline
    \hline
    Th:X & $\eta_\textrm{escape}$ (\%) & Th:X & $\eta_\textrm{escape}$ (\%) \\
    \hline
    Th:Ag & 8  &Th:Nb & 21  \\
    Th:Al & 9  &Th:Ni & 7  \\
    Th:Au & 4  &Th:Pd & 6  \\
    Th:Ba  & 20  &Th:Pt & 3  \\
    Th:Cu & 5  &Th:Ti & 18  \\
    Th:Hf & 35  &Th:V & 20  \\
    Th:Lu & 29  &Th:Zr & 23  \\
    Th:Mo & 9  & & \\
    \hline
    \hline
    \end{tabular}
    \caption{DFT aided simulations of $\eta_\textrm{escape}$ for a number of host materials for implantation at 100 eV.}\label{tab:escape}
\end{table}

\section{Crystal Characterisation}
\label{sec:channel}
\begin{figure}[!htbp]
\begin{center}
\includegraphics[width=\columnwidth]{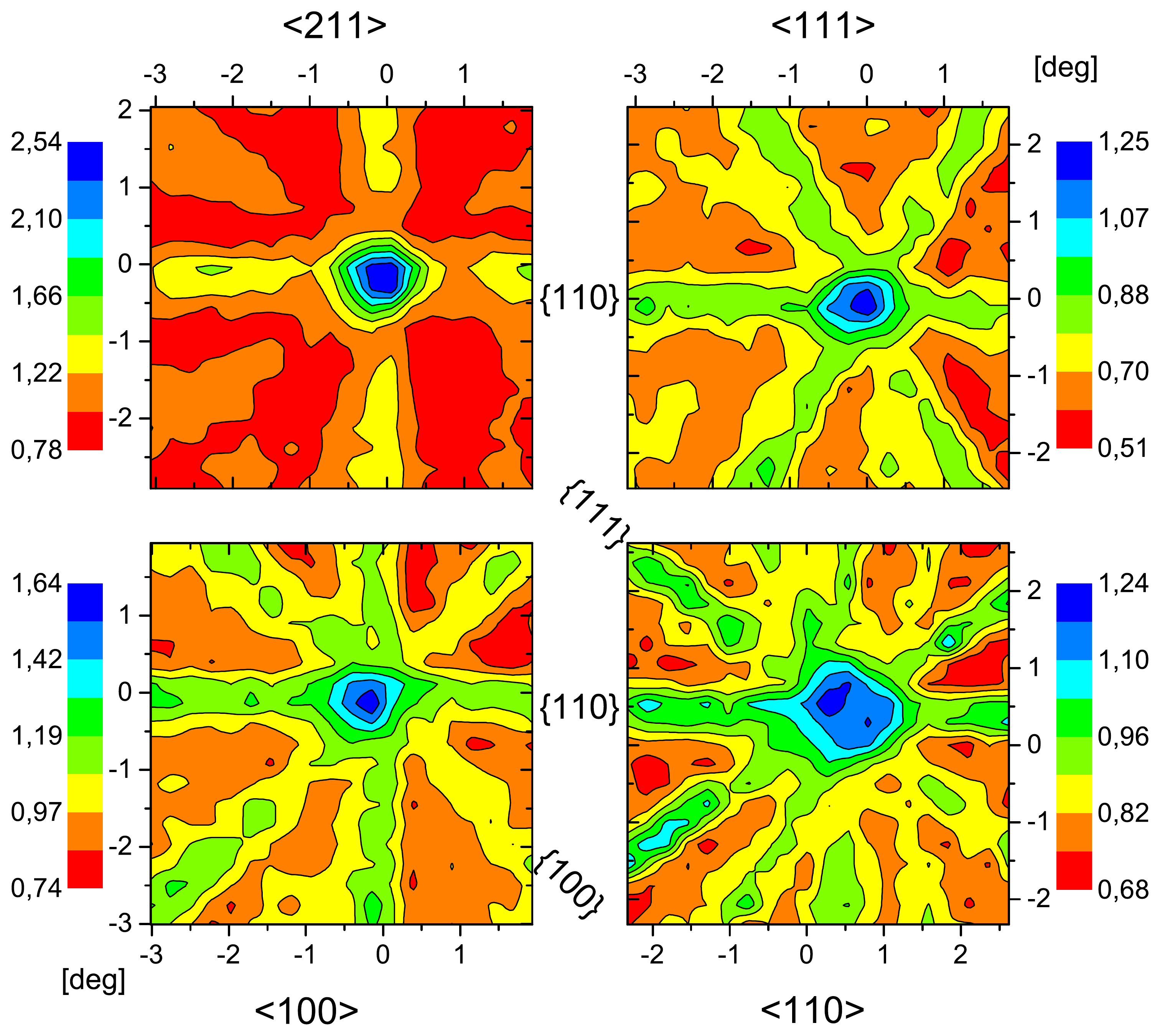}
\caption{Experimental 2D emission channeling patterns of $^{229}$Ac implanted at room temperature into CaF$_2$, measured in the vicinity ($\pm$ $2$ deg) of four major crystallographic directions ($\left\langle 211\right\rangle $, $\left\langle 111\right\rangle $,$\left\langle 100\right\rangle $ and $\left\langle 110\right\rangle $). The color scale represents the relative amount of electrons hitting the detector, where blue zones indicate that more electrons are emitted. The green lines represent crystallographic planes along which the electrons channel.}
\label{fig:4} 
\end{center}      
\end{figure}

To detect the radiative decay of the $^{229\textrm{m}}$Th isomer in the VUV spectroscopy experiment proposed in section \ref{sec:vuv}, it is essential to inhibit the IC decay channel. As mentioned above, this can be achieved by using large band gap crystals as a host for the $^{229}$Th isomer. Due to the band gap being larger than the isomer energy, there are no available electronic states for internal conversion and the crystal, without dopants or colour centers, is transparent to the emitted VUV photon (or incoming VUV photons) upon decay to the $^{229}$Th ground state. However, depending on the local atomic configuration (e.g. occupied lattice site and charge compensation mechanism) of the thorium impurities in the host crystal, states can be introduced inside the band gap energy region. If these gap states reduce the size of the band gap below the isomer energy, the IC decay channel is no longer suppressed and will dominate over the radiative channel. In addition, a hyperfine structure originating from a non-vanishing electric field gradient at the Th site causes a homogeneous shift in transition energy of the order of $10^{\textrm{-}5}$ eV \cite{Kazakov2012}. Since this shift also depends on the local configuration it is necessary for all thorium atoms to be in the same lattice location and configuration for a final VUV spectroscopy experiment aiming at a highest accuracy. \\
\medskip
\\The concept to suppress the IC decay channel with a large band gap material is illustrated for CaF$_2$, a well-studied material and a suited host crystal with a direct band gap of $12.1$ eV \cite{Rubloff1972}. Theoretical calculations predict a lowest-energy configuration of Th atoms in a CaF$_2$ crystal, whereby the Th$^{4+}$ ions occupy Ca$^{2+}$ substitutional sites with a charge compensation mechanism of two neighboring F$^{-}$ interstitials \cite{Dessovic2014}. Whereas this local configuration is not expected to reduce the band gap significantly, allowing for high-resolution spectroscopy, there is no experimental insight in the occupancy fraction of this configuration, depending on the production method of the $^{229\textrm{m}}$Th ensemble. A low occupancy fraction could present a severe loss in signal during VUV spectroscopy. In the context of the spectroscopy approach proposed in this work, it is crucial to identify and quantify eventual non-substitutional $^{229}$Th incorporation resulting from the implantation of a $^{229}$Ac parent isotope (e.g. in interstitial or disordered sites). The emission channeling technique is particularly suitable for such studies \cite{Hofsass1991}. After the implantation of a suitable radioactive isotope, the particles emitted upon decay ($\beta^{-}$, $\beta^{+}$ or $\alpha$) interact with the screened periodic Coulomb potential of the crystal lattice. The atomic rows and planes of the crystal determine an anisotropic scattering of the emitted particles, resulting in emission patterns that are very sensitive to the exact position of the radioactively decaying nucleus within the lattice. A sensitivity of $\approx 0.1$ $\textrm{\AA}$ can be reached for the position of the radioactive species \cite{Pereira2014}. Since the substitutional thorium configuration is the lowest-energy configuration it might be possible to increase its default occupancy fraction by annealing. A position-sensitive detector measures the 2D electron emission patterns around selected crystal axes. Fitting these experimental data with linear combinations of patterns simulated for various possible lattice sites allows to quantitatively and unambiguously determine the lattice site occupancy. Lattice sites with an occupancy fraction down to $5$\% are detectable in this way.\\
\medskip
\\At ISOLDE, CERN, multiple actinium isotopes can be implanted into CaF$_2$ to determine the lattice location of thorium dopants. The $^{229}$Th isotope can not be studied directly due to its very long half-life (T$_{1/2} \approx 8000 \textrm{ y}$, $\alpha$ emitter). Nevertheless, emission channeling experiments can be performed on implanted $^{229}$Ac ($T_{1/2} = 62.7\textrm{ min}$, $\beta^{-}$ emitter) to determine the lattice location of $^{229\textrm{m}}$Th's mother isotope. An implantation yield of $10^6$ pps is required in this case. The $2.3$ eV recoil energy transferred to the $^{229}$Th daughter nucleus upon decay is not sufficient for atomic displacement. Therefore, the $^{229}$Th atoms are expected to occupy the same lattice site as the implanted $^{229}$Ac atoms. A first series of emission channeling experiments with $^{229}$Ac has been performed in the online EC-SLI set-up at ISOLDE, CERN \cite{Silva2013}. The measured 2D $\beta^-$ emission channeling patterns from $^{229}$Ac, implanted at room temperature, in the vicinity of the CaF$_2$ $\langle 211\rangle$, $\langle 111\rangle$, $\langle 100\rangle$ and $\langle 110\rangle$ crystallographic directions are shown in \cref{fig:4}. The symmetry of these patterns corresponds to the crystal structure of CaF$_2$. This indicates that the CaF$_2$ crystal does not sustain severe damage under this heavy ion implantation which could hamper emission channeling experiments. Additionally, the data show that a large fraction of the $^{229}$Ac atoms occupy calcium substitutional positions. Further analysis of these patterns is ongoing.\\
\medskip
\\Besides $^{229}$Ac, other isotopes can be used as substitute. Different isotopes of the same element are chemically identical. Hence, the local atomic configuration of thorium and actinium atoms in a crystal should not depend on the implanted isotope itself. In the future, it could be possible for the laser ion source at ISOLDE to also produce a ${}^{231}$Ac (T$_{1/2} = 7.5 \textrm{ min}$) beam with yields of $10^6$ pps. $^{231}$Ac is an exceptionally well-suited parent isotope for emission channeling experiments because it decays ($\beta^{-}$) to ${}^{231}$Th (T$_{1/2} = 25.5 \textrm{ h}$), which, unlike ${}^{229}$Th, also decays via $\beta^{-}$. The significant difference in lifetime allows the study of the lattice incorporation of the ${}^{231}$Ac parent (online) and, afterwards, of the ${}^{231}$Th daughter (offline)  \cite{Wahl2004}. These two successive  $\beta^{-}$ decays after the implantation of ${}^{231}$Ac is an ideal proxy for what occurs with the ${}^{229}$Ac/${}^{229}$Th decay in the VUV spectroscopy experiment proposed here, since the positions of both ${}^{231}$Ac and ${}^{231}$Th can be studied in the same experiment, separately.

\section{Summary}
\noindent
To allow for the development of the $^{229\textrm{m}}$Th's potential applications, two key observables remain unknown: its radiative lifetime and excitation energy. The way the isomer is produced has a distinct influence on the results, when attempting to determine these observables through the detection of the radiative $^{229\textrm{m}}$Th-$^{229}$Th transition. To this end, a concept to use $^{229}$Ac's $\beta$ decay as an alternative production method of the $^{229\textrm{m}}$Th isomer is presented. The photons of the $^{229\textrm{m}}$Th-$^{229}$Th transition will be identified and scrutinized through a high-resolution spectrometer after feeding of the isomer with the decay of implanted $^{229}$Ac ions in a large-bandgap host. Production of $^{229\textrm{m}}$Th via $^{229}$Ac reveals a number of benefits over the commonly used $\alpha$ decay of $^{233}$U. These advantages include the availability of an isotopically pure, intense source of $^{229}$Ac from the ISOLDE facility, an expected larger $\beta$ branching ratio towards $^{229\textrm{m}}$Th, a more manageable decay half-life, a lower radioluminescence background for VUV spectrometery compared to other concepts and a controlled background characterization. Preparatory experiments, testing key ingredients of this concept, were conducted at ISOLDE. These experiments have confirmed, first, the availability of a high intensity, pure $^{229}$Ac ion beam. Next, in an attempt to determine the $\beta$-decay branching ratio towards the isomer, no signal with its anticipated 7 $\mu$s half-life was observed and further analysis should provide more insight. Finally, emission channeling patterns were obtained for $^{229}$Ac ions, implanted at 30 keV energy in a CaF$_2$ crystal, indicating a significant calcium substitutional positioning in the CaF$_2$ host.

\begin{acknowledgments}
\noindent
This work has received funding from Research Foundation Flanders (FWO, Belgium), by GOA/2015/010, C14/18/074 (BOF KU Leuven), from the European Union’s Horizon 2020 research and innovation programme under grant agreement No 654002 (ENSAR2) and No824096 (RADIATE), from the European Research Council under the ERC-2011-AdG-291561-HELIOS, from the Spanish MINECO via project FPA-2015-65035-P and from the FCT-Portugal, project CERN-FIS-PAR-0005-2017.
\end{acknowledgments}

\bibliography{229Acpaper_references}

\end{document}